\documentclass[prb,letterpaper,aps,floatfix,twocolumn]{revtex4-1}
\usepackage{graphicx}
\usepackage{amsmath}
\usepackage{slashed}
\newcommand{\beq}{\begin{equation}}
\newcommand{\eeq}{\end{equation}}
\newcommand{\bk}{{{\bf{k}}}}

\newcommand{\br}{{{\bf{r}}}}

\newcommand{\bP}{{{\bf{P}}}}

\newcommand{\bA}{{\bf{A}}}
\newcommand{\bB}{{\bf{B}}}
\newcommand{\bM}{{\bf{M}}}
\newcommand{\bE}{{\bf{E}}}

\newcommand{\bj}{{\bf{j}}}

\newcommand{\bq}{{\bf{q}}}

\newcommand{\bb}{{\bf{b}}}

\newcommand{\beqa}{\begin{eqnarray}}
\newcommand{\eeqa}{\end{eqnarray}}

\newcommand{\dg}{{\dag}}
\newcommand{\pdg}{{\vphantom \dag}}
\newcommand{\btau}{{\boldsymbol \tau}}
\newcommand{\bsigma}{{\boldsymbol \sigma}}

\newcommand{\bnabla}{{\boldsymbol \nabla}}
\newcommand{\bpi}{{\boldsymbol \pi}}
\begin{document}
\title{Topological response in Weyl semimetals and the chiral anomaly}
\author{A.A. Zyuzin}
\author{A.A. Burkov}
\affiliation{Department of Physics and Astronomy, University of Waterloo, Waterloo, Ontario 
N2L 3G1, Canada}
\date{\today}
\begin{abstract}
We demonstrate that topological transport phenomena, characteristic of Weyl semimetals, namely the semi-quantized anomalous
Hall effect and the chiral magnetic effect (equilibrium magnetic-field-driven current), may be thought of as two distinct manifestations of the 
same underlying phenomenon, the chiral anomaly. We show that the topological response in Weyl semimetals is fully described by a 
$\theta$-term in the action for the electromagnetic field, where $\theta$ is not a constant parameter, like e.g. in topological insulators, 
but is a field, which has a linear dependence on the space-time coordinates. 
We also show that the $\theta$-term and the corresponding topological response survive for sufficiently weak translational symmetry breaking perturbations, 
which open a gap in the spectrum of the Weyl semimetal, eliminating the Weyl nodes. 
\end{abstract}
\maketitle
\section{Introduction}
\label{sec:1} 
Weyl semimetals have attracted attention recently as a new kind of topologically-nontrivial phase of matter: Weyl semimetal 
is gapless in the bulk yet possesses protected surface states and the corresponding topological transport 
phenomena.~\cite{Vishwanath11,Balents11,Ran11,Burkov11-1,Burkov11-2,Burkov12-1,Kim11,Fang11,Halasz11,Hosur11,Aji11,Carpentier12,Son12,Qi12,Grushin12,Balents12,Kolomeisky12,Garate12,Jiang12}
Topological protection in this case results from the separation of the individual Weyl band-touching nodes with opposite topological charges in momentum space, which makes it impossible 
to hybridize the nodes and produce a fully gapped insulating state without violating translational symmetry.~\cite{Volovik03,Volovik05,Volovik07,Volovik11}
Such separation requires breaking of either time-reversal (TR) or inversion (I) symmetry, or both,~\cite{Murakami} as in the presence of TR and I all bands are 
doubly-degenerate by Kramers theorem. 

As has long been known in the quantum field theory context, chiral Weyl fermions are associated with the phenomenon 
of chiral anomaly.~\cite{Adler69,Jackiw69,Nielsen83}  
Chiral anomaly manifests in nonconservation of the numbers of particles of a specific chirality in the presence of topologically-nontrivial 
configurations of the background gauge field (electromagnetic field in our context), even though these numbers are conserved classically (for massless
particles). This phenomenon plays an important role in the Standard model of particle physics.~\cite{Fujikawa,Nakahara}
In the condensed matter context, the 2+1-dimensional relative of the chiral anomaly, the parity anomaly, has mainly been discussed, due to its 
close relation to the quantum Hall effect.~\cite{Jackiw84,Fradkin86,Semenoff84,Haldane88,Ludwig94}
The discovery of Weyl semimetals provides a concrete condensed matter system, where 3+1-dimensional chiral anomaly and related effects can be 
realized.~\cite{Aji11,Son12,Qi12}  

In this paper, we focus on a specific realization of a Weyl semimetal in a magnetically-doped multilayer heterostructure, made of alternating 
layers of topological insulator~\cite{Kane,Zhang} (TI) and normal insulator (NI) materials.~\cite{Burkov11-1}
This system realizes the simplest possible kind of Weyl semimetal, with only two Weyl nodes of opposite chirality, the smallest 
number allowed by the Nielsen-Ninomiya theorem,~\cite{Nielsen81} in its bandstructure (identical results are obtained by magnetically doping a 
bulk TI with a small bandgap). 
We have demonstrated before that such a system possesses topologically-nontrivial transport properties, 
namely the semi-quantized anomalous Hall effect~\cite{Burkov11-1,Burkov11-2} and the chiral magnetic effect (generation of equilibrium current by magnetic field).~\cite{Burkov12-1}
The chiral magnetic effect has been known for 
some time in the particle physics context~\cite{Vilenkin80,Cheianov98,Kharzeev08,Kharzeev11,Son12} and may have recently been observed experimentally in 
relativistic heavy ion collisions.~\cite{Abelev09}
Observation of this effect in Weyl semimetals would be of significant interest. 

In this work we demonstrate that both the quantum anomalous Hall effect and the chiral magnetic effect in Weyl semimetals are manifestations of the same 
underlying phenomenon, the chiral anomaly. We show that opposite-chirality Weyl nodes, separated in momentum space and in energy, 
give rise to an induced $\theta$-term in the action of the electromagnetic field
\beq
\label{eq:1}
S_{\theta} = \frac{e^2}{32 \pi^2} \int d t d \br \,\,\theta(\br, t)  \epsilon^{\mu \nu \alpha \beta} F_{\mu \nu} F_{\alpha \beta},
\eeq
where $\hbar = c =1$ units are used henceforth. 
$\theta(\br, t)$ is an ``axion" field,~\cite{Wilczek87} which has the following form
\beq
\label{eq:2}
\theta(\br , t) =  2 \bb \cdot \br - 2 b_0 t, 
\eeq
where $2 \bb$ is the separation between the Weyl nodes in momentum space and $2 b_0$ is the separation between the nodes in energy. 

In the rest of the paper, we will give a derivation of Eq.~\eqref{eq:1} using Fujikawa's method,~\cite{Fujikawa79,Fujikawa,Nakahara} which clearly demonstrates the relation of the $\theta$-term 
to the chiral anomaly, and show that both the anomalous Hall and the chiral magnetic effects follow directly from Eq.~\eqref{eq:1}. 
We will also demonstrate that, somewhat contrary to the commonly expressed belief that Weyl semimetal is only topologically stable in the presence of translational symmetry, 
which prohibits the mixing of Weyl nodes, the $\theta$-term in Eq.~\eqref{eq:1} in fact survives even when the translational symmetry is broken and the Weyl nodes are hybridized 
and gapped out, provided the translational symmetry breaking is sufficiently weak.The quantum anomalous Hall effect and the chiral magnetic effect are thus more robust than the Weyl nodes 
themselves. 

\section{$\theta$-term in Weyl semimetals}
\label{sec:2}
We start from a specific realization of a Weyl semimetal in a TI-NI multilayer heterostructure.~\cite{Burkov11-1}
The advantage of this system is its simplicity (and, perhaps, simplicity of experimental realization as well), as the Weyl semimetal realized in this system contains only two Weyl nodes in its bandstructure, 
i.e. the minimal number, required by the fermion-doubling theorem.~\cite{Nielsen81} 
The Hamiltonian, describing the multilayer system, is given by
\beqa
\label{eq:3}
H&=&\sum_{\bk_{\perp}, ij} \left[ v_F \tau^z (\hat z \times \bsigma) \cdot \bk_{\perp} \delta_{i,j} + m \sigma^z \delta_{i,j}
+ \Delta_S \tau^x \delta_{i,j} \right. \nonumber \\ 
&+& \left.\frac{1}{2} \Delta_D \tau^+ \delta_{j, i+1} + \frac{1}{2} \Delta_D \tau^- \delta_{j, i-1} \right] c^\dg_{\bk_{\perp} i} c^\pdg_{\bk_{\perp} j}. 
\eeqa
The first term in Eq.(\ref{eq:3}) describes the two (top and bottom) surface states of an individual TI layer. 
$v_F$ is the Fermi velocity, characterizing the surface Dirac fermion, which we take to be the same on the top and 
bottom surfaces of each layer. $\bk_{\perp}$ is the momentum in the 2D surface Brillouin zone (BZ).
$\bsigma$ is the triplet of Pauli matrices, acting on the real spin degree of freedom, and $\btau$ are Pauli matrices, acting 
on the {\em which surface} pseudospin degree of freedom. The indices $i,j$ label distinct TI layers. 
The second term describes exchange spin splitting of the surface states, which is induced by doping 
the sample with magnetic impurities.
The remaining terms in Eq.(\ref{eq:3}) describe tunneling between top and bottom surfaces within the same TI layer (the term, proportional 
to $\Delta_S$), and between top and bottom surfaces of neighboring TI layers (terms, proportional to $\Delta_D$).  
Diagonalizing Eq.~\eqref{eq:3} one finds, when $(\Delta_S - \Delta_D)^2 < m^2 < (\Delta_S + \Delta_D)^2$, two Weyl nodes, separated 
along the $z$-axis in momentum space by a wavevector of magnitude
\beq
\label{eq:4}
2 b  = \frac{2}{d}\textrm{arccos}\left(\frac{\Delta_S^2 + \Delta_D^2 - m^2}{2 \Delta_S \Delta_D} \right), 
\eeq
where $d$ is the multilayer period. 

The multilayer model Eq.~\eqref{eq:3} possesses inversion symmetry with 
respect to an inversion center, placed midway between the surfaces of any TI or NI layer. This symmetry guarantees that the two 
Weyl nodes occur at the same energy as the corresponding symmetry operation interchanges the nodes with opposite chirality. 
In a real multilayer, this symmetry will likely not be present, and can in fact also be removed on purpose by making the top and bottom 
surfaces in each TI layer distinct by, e.g. creating a potential drop between them. In the absence of the inversion symmetry, the 
Weyl nodes of opposite chirality will then be shifted with respect to each other in energy, as well as in momentum.~\cite{Burkov12-1}
This is achieved by a spin-orbit interaction term $\lambda \tau^y \sigma^z$, which is allowed by the broken inversion symmetry. 
Adding this term to the multilayer Hamiltonian Eq.~\eqref{eq:3}, leads to the energy separation between the Weyl nodes 
of magnitude~\cite{Burkov12-1}
\beq
\label{eq:5}
2 b_0 = \frac{\lambda}{\Delta_D m} \sqrt{(m_{c2}^2 - m^2) (m^2 - m_{c1}^2)}, 
\eeq
where $m_{c1}^2 = (\Delta_S - \Delta_D)^2$ and $m_{c2}^2 = (\Delta_S + \Delta_D)^2$. 

To generalize and simplify the subsequent considerations, we will move away from the microscopic model 
of TI-NI multilayer Eq.~\eqref{eq:3} and introduce a corresponding low-energy model, obtained by expanding the   
microscopic multilayer Hamiltonian around the location of the Weyl nodes. 
The justification for this is that we want to describe universal phenomena, which depend only on the number and energy-momentum 
separation between the Weyl nodes of opposite chirality, but not on any other details of the energy spectrum away from the 
Weyl nodes. 
It can be demonstrated by an explicit calculation,~\cite{Burkov11-1,Burkov12-1} that the properties of interest to us
do not change if a fully microscopic model is used. 
 
We then obtain a generic low-energy model 
of two Weyl nodes of opposite chirality, separated in momentum space and in energy, described by the momentum-space Hamiltonian
\beq
\label{eq:6}
H = \tau^z \bsigma \cdot \bk + \tau^z b_0 + \bsigma \cdot \bb, 
\eeq
where we have absorbed the Fermi velocity (in general different in different directions) in the definition of momentum. 
The operators $\btau$ and $\bsigma$ now have a meaning, different from Eq.~\eqref{eq:3}. $\btau$ now describes the 
node degree of freedom, while $\bsigma$ describes the conduction-valence band degree of freedom (nondegenerate conduction and valence bands touch at the 
Weyl nodes). 
Finally, we couple the electrons to an external electromagnetic field, and represent the system in terms of the 
imaginary time action
\beqa
\label{eq:7}
S&=&\int d \tau d \br \,\,\psi^\dg \left[ \partial_{\tau} + i e A_0 + b_0 \tau^z\right. \nonumber \\ 
&+&\left.\tau^z \bsigma \cdot \left( -i \bnabla + e \bA + \bb \tau^z \right) \right] \psi^\pdg,
\eeqa
where $A_{\mu} = (A_0, \bA)$ is the electromagnetic gauge potential and $\psi^\dg, \psi^\pdg$ are the 4-component spinor Grassman field variables. We have suppressed all 
explicit spinor indices in the Grassmann variables for brevity. 
We now make the following observation, that will play a crucial role in our analysis. The imaginary time action Eq.~\eqref{eq:7} possesses a chiral symmetry
\beq 
\label{eq:8}
\psi \rightarrow e^{-i \tau^z \theta/2} \psi, 
\eeq 
which expresses an apparent separate conservation of the number of fermions of left and right chirality. This suggests that the terms $\tau^z b_0$ and $\bsigma \cdot \bb$ in Eq.~\eqref{eq:7}
can be eliminated by a gauge transformation:
\beq
\label{eq:9}
\psi \rightarrow e^{- i \tau^z \theta(\br, \tau)/2} \psi,\,\, \psi^\dg \rightarrow \psi^\dg e^{i \tau^z \theta(\br, \tau)/2},
\eeq
where $\theta(\br, \tau) = 2 \bb \cdot \br - 2 i b_0 \tau$ and one should keep in mind that $\psi$ and $\psi^\dg$ are not complex conjugates of each other, but are independent variables in the 
fermion path integral. 
The imaginary time action then becomes
\beq
\label{eq:10}
S = \int d \tau d \br \,\, \psi^\dg \left[ \partial_{\tau} + i e A_0 
 + \tau^z \bsigma \cdot \left( -i \bnabla + e \bA \right) \right] \psi^\pdg, 
\eeq
which describes two Weyl nodes of opposite chirality, existing at the same point in momentum space and in energy. 
This argument then leads one to the conclusion that the system of Weyl nodes, separated in energy and momentum, is equivalent 
to the system of two degenerate Weyl nodes and thus does not possess any special transport properties, which we know is incorrect. 
The missing link in the above naive argument is precisely the chiral anomaly: while the imaginary time action Eq.~\eqref{eq:7} does indeed 
possess the chiral symmetry, the gauge transformation of Eq.~\eqref{eq:9} changes not only the action itself, but also the measure of the path 
integral, representing the partition function of the system
\beq
\label{eq:11}
Z = \int D \psi^\dg D \psi e^{-S[\psi^\dg,\psi]}, 
\eeq
where we assume, for simplicity, that the electromagnetic field does not fluctuate (our results do not depend on this assumption, as will be clear from the derivation below). 
As we will demonstrate, the change in the path integral measure, induced by the chiral gauge transformation Eq.~\eqref{eq:9}, gives rise precisely to the additional $\theta$-term 
in the action, given, after Wick rotation to real time, by Eq.~\eqref{eq:1}.

To derive the $\theta$-term we will use a simple modification of the Fujikawa's method,~\cite{Fujikawa79,Fujikawa,Nakahara} used originally to derive the chiral anomaly in the path integral language. 
We will provide all details of the derivation for readers which may not be familiar with the method. 
For a related application of this method to 3D TI see Ref.~\onlinecite{Vishwanath10}.

To begin, it is convenient to rewrite the imaginary time action Eq.~\eqref{eq:7} in the standard ``relativistic" notation. 
We introduce Dirac $\gamma$-matrices as 
\beq
\label{eq:12}
\gamma^0 = \tau^x,\,\, \gamma^{\mu} = i \tau^y \sigma^{\mu}, \,\, \gamma^5 = - i \gamma^0 \gamma^1 \gamma^2 \gamma^3 = \tau^z. 
\eeq
Further defining $\gamma^4 = - i \gamma^0$, we can rewrite Eq.~\eqref{eq:7} as
\beq
\label{eq:13}
S = \int d^4 x \,\, \bar \psi\, i \gamma^{\mu} \left(\partial_{\mu} + i e A_{\mu} + i b_{\mu} \gamma^5 \right) \psi, 
\eeq
where $\mu = 1,\ldots,4$, $\bar \psi = \psi^\dg \gamma^0$, $b_4 = - i b_0$, and $\int d^4 x$ denotes integral over the 4-dimensional Euclidean space-time. 
We note that all matrices $\gamma^{\mu}$ are antihermitian, while $\gamma^5$ is hermitian, and $\gamma^5$ anticommutes with $\gamma^{\mu}$. 

To eliminate the $b_{\mu}$ term from Eq.~\eqref{eq:13} we will apply an infinite sequence of infinitesimal chiral gauge transformations
\beqa
\label{eq:14}
&&\psi \rightarrow e^{- i d s \,\theta(x) \gamma^5/2} \psi, \nonumber \\
&&\bar \psi \rightarrow \bar \psi e^{- i  d s  \, \theta(x) \gamma^5/2},
\eeqa
where $\theta(x) = 2 b_{\mu} x_{\mu}$ and the sign of the exponential in the second line above follows from the fact that $\gamma^5$ anticommutes with $\gamma^0$. 
The variable $s \in [0,1]$, whose differential $d s$ appears in the infinitesimal gauge transformation above, parametrizes the infinite sequence of the chiral gauge 
transformations of Eq.~\eqref{eq:14}. 

We need to find how the infinitesimal transformation Eq.~\eqref{eq:14} changes the path integral measure $D \bar \psi D \psi$. Following Fujikawa,~\cite{Fujikawa79}
we consider the 3+1 dimensional Dirac operator, taken at stage $s$ of the sequence of infinitesimal chiral transformations
\beq
\label{eq:15}
\slashed{D} = \gamma^{\mu} [\partial_{\mu} + i e A_{\mu} + i b_{\mu} (1-s) \gamma^5]. 
\eeq
Since $\gamma^{\mu}$ are antihermitian, $\slashed{D}$ is a hermitian operator.   
Suppose we can solve the eigenvalue problem for the hermitian operator $\slashed{D}$
\beq
\label{eq:16}
\slashed{D} \phi_n(x) = \epsilon_n \phi_n(x), 
\eeq
where $\epsilon_n$ are real eigenvalues and $\phi_n(x)$ are 4-component spinor eigenfunctions (we suppress the spinor indices for brevity). 
We assume that $\phi_n(x)$ wavefunctions can be normalized to unity (this requires assuming a finite space-time volume and taking the volume to infinity at the end)
\beq
\label{eq:17}
\int d^4 x \, \phi^*_n(x) \phi^\pdg_m(x) = \delta_{nm}. 
\eeq
Given a complete set of eigenfunctions $\phi_n(x)$, we can expand the Grassmann variables $\psi$ and $\bar \psi$ in the path integral as
\beq
\label{eq:18}
\psi(x) = \sum_n \phi_n(x) c_n,\,\, \bar \psi(x) = \sum_n \phi^*_n(x) \bar c_n, 
\eeq 
where $c_n$ and $\bar c_n$ are the new Grassmann variables. 
Analogously, we can expand the transformed Grassmann fields
\beqa
\label{eq:19}
&&\psi'(x) = e^{- i ds  \,\theta(x) \gamma^5/2} \psi(x) \nonumber \\
&=&[1 - i ds \, \theta(x) \gamma^5/2] \sum_n \phi_n(x) c_n  = \sum_n \phi_n(x) c'_n, \nonumber \\
&&\bar \psi'(x) = \bar \psi(x) e^{- i ds \, \theta(x) \gamma^5/2} \nonumber \\
&=&\sum_n \phi^*_n(x) \bar c_n [1 - i d s  \, \theta(x) \gamma^5/2] = \sum_n \phi^*_n(x) \bar c'_n. \nonumber \\
\eeqa
Defining the infinitesimal chiral transformation operator as
\beq
\label{eq:20}
U_{nm} = \delta_{nm} - ds  \frac{i}{2} \int d^4 x \, \phi_n^*(x) \theta(x) \gamma^5 \phi_m(x), 
\eeq
we obtain
\beq
\label{eq:21}
c'_n = \sum_{m} U_{nm} c_m, \,\, \bar c'_n = \sum_m U_{mn} \bar c_m.
\eeq
This immediately gives us the path integral Jacobian, corresponding to the chiral gauge transformation 
\beqa
\label{eq:22}
J&=&\textrm{det}(U^{-2}) = e^{\ln \textrm{det} (U^{-2})} = e^{-2 \textrm{Tr} \ln (U)} \nonumber \\
&=&e^{ i ds  \int d^4 x \sum_n \phi^*_n(x) \theta(x) \gamma^5 \phi_n(x)}. 
\eeqa
Consider the quantity, appearing in the exponential in Eq.~\eqref{eq:22}
\beq
\label{eq:23}
I(x) = \sum_n \phi_n^*(x) \gamma^5 \phi_n(x). 
\eeq
To understand the meaning of this quantity, we note that $\gamma^5$ anticommutes with the Dirac operator $\slashed{D}$. 
This means that if $\phi_n(x)$ is an eigenfunction of $\slashed{D}$ with an eigenvalue $\epsilon_n$, then $\gamma^5 \phi_n(x)$ 
is an eigenfunction of $\slashed{D}$ with eigenvalue $-\epsilon_n$. 
It follows that since the eigenvectors of a hermitian operator $\slashed{D}$, corresponding to nondegenerate eigenvalues, are orthogonal, 
only zero eigenmodes contribute to $\int d^4 x I(x)$. Then we obtain
\beq
\label{eq:24}
\int d^4 x \, I(x) = n_+ - n_- = \textrm{ind}(\slashed{D}), 
\eeq
where $n_{\pm}$ is the number of zero-mode eigenstates with positive (negative) eigenvalue of $\gamma^5$ (i.e. chirality). 
Thus, with a slight abuse of terminology, we can call $I(x)$ a ``local index" or ``index density" of the Dirac operator $\slashed{D}$, in the sense that the integral of $I(x)$ 
over the 3+1-dimensional space-time gives the {\em analytical index} of $\slashed{D}$. 

To evaluate $I(x)$ explicitly we use the standard method of heat kernel regularization.~\cite{Nakahara} 
The regularization is necessary, because, as written in Eq.~\eqref{eq:23}, $I(x)$ is poorly defined, since a finite result is obtained due to mutual cancellation 
of divergent contributions. 
We have
\beqa
\label{eq:25}
I(x)&=&\lim_{M \rightarrow \infty} \sum_n \phi^*_n(x) \gamma^5 e^{-\epsilon_n^2/M^2} \phi_n(x) \nonumber \\
&=&\lim_{M \rightarrow \infty} \sum_n \phi^*_n(x) \gamma^5 e^{-\slashed{D}^2/M^2} \phi_n(x). 
\eeqa
The square of the Dirac operator in the exponential is given by
\beqa
\label{eq:26}
&&\slashed{D}^2 =- D_{\mu} D_{\mu} - (1-s)^2 b_{\mu} b_{\mu} + \frac{i e}{4} [\gamma^{\mu}, \gamma^{\nu}] F_{\mu \nu} \nonumber \\ 
&+&i (1-s) [\gamma^{\mu}, \gamma^{\nu}] b_{\mu} D_{\nu} \gamma^5, \nonumber \\
\eeqa
where $D_{\mu} \equiv \partial_{\mu} + i e A_{\mu}$ and we have used
\beq
\label{eq:27}
[D_{\mu}, D_{\nu}] = i e (\partial_{\mu} A_{\nu} - \partial_{\nu} A_{\mu}) = i e F_{\mu \nu}.
\eeq
Substituting this into Eq.~\eqref{eq:25}, and using the completeness relation
\beq
\label{eq:28}
\sum_n \phi_n^*(x) \phi_n(y) = \delta(x - y), 
\eeq
we obtain
\beqa
\label{eq:29}
&&I(x) = \lim_{M \rightarrow \infty} \int \frac{d^4 k}{(2 \pi)^4} \textrm{tr} \gamma^5 e^{-i k x} e^{-\slashed{D}^2/M^2} e^{i k x} \nonumber \\
&=&\lim_{M \rightarrow \infty} \int \frac{d^4 k}{(2 \pi)^4} \textrm{tr} \gamma^5 \exp \left[ \frac{(i k_{\mu} + D_{\mu})^2}{M^2} + \frac{(1-s)^2 b_{\mu}b_{\mu}}{M^2}\right. \nonumber \\
&-&\left.\frac{i e }{4 M^2} [\gamma^{\mu}, \gamma^{\nu}] F_{\mu \nu} - \frac{i (1-s)}{M^2} [\gamma^{\mu}, \gamma^{\nu}] b_{\mu} (i k_{\nu} + D_{\nu}) \gamma^5 \right]. \nonumber \\
\eeqa
Rescaling momentum integration variable $k_{\mu} \rightarrow M k_{\mu}$ and leaving only terms that survive the limit $M  \rightarrow \infty$ and the trace operation (which must contain 
four $\gamma$-matrices and be proportional to $1/M^4$), we finally obtain
\beqa
\label{eq:30}
I(x)&=&-\frac{e^2}{32} \textrm{tr} \gamma^5 [\gamma^{\mu},\gamma^{\nu}] [\gamma^{\alpha}, \gamma^{\beta}] F_{\mu \nu} F_{\alpha \beta} \nonumber \\
&=&\frac{e^2}{32 \pi^2} \epsilon^{\mu \nu \alpha \beta} F_{\mu \nu} F_{\alpha \beta}.
\eeqa
As pointed out by Fujikawa,~\cite{Fujikawa79} Eqs.~\eqref{eq:24} and~\eqref{eq:30} can be thought of as  a local version of the Atiyah-Singer index theorem.~\cite{Nakahara} 
When integrated over space-time, Eq.~\eqref{eq:30}
connects the analytical index of the Dirac operator $\slashed{D}$ with its topological index.~\cite{Franz10}

Substituting Eq.~\eqref{eq:30} back into the expression for the Jacobian of the infinitesimal chiral gauge transformation at ``time" $s$, we obtain
\beq
\label{eq:31}
J = e^{- i ds \int d^4 x \theta(x) I(x)}. 
\eeq
To get the total contribution to the action from the Jacobian after the $b_{\mu} \gamma^5$ term has been fully eliminated from Eq.~\eqref{eq:13} we
integrate over the variable $s$
\beqa
\label{eq:32}
S_{\theta}&=&  i \int_0^1 ds  \int d^4 x \, \theta(x) I(x) \nonumber \\
&=&\frac{i e^2}{32 \pi^2} \int d^4 x  \,\theta(x) \epsilon^{\mu \nu \alpha \beta} F_{\mu \nu} F_{\alpha \beta}. 
\eeqa
The dependence of the imaginary time action of the system on $b_{\mu}$ has thus been fully transferred to $S_{\theta}$, which describes completely 
the topological electromagnetic response of Weyl semimetal. 
After analytical continuation to real time $\tau \rightarrow  i t$, we obtain Eq.~\eqref{eq:1}. 
Topological response of Weyl semimetals is thus described by an axion-type action, with the ``axion field" $\theta(\br, t)$, which depends linearly 
on the space-time coordinates. It is useful to compare this with the $\theta$-term in the action of the electromagnetic field, characteristic of TIs. 
In that case $\theta = \pi$, which is the only nonzero value, consistent with TR symmetry. The type of Weyl semimetal we are considering in this paper 
can be thought of as being obtained from a TI in which both TR and I symmetries have been broken.~\cite{Burkov11-1,Burkov12-1}
It is easy to see that broken TR allows for a nontrivial dependence of $\theta$ on the spatial coordinates, while broken I allows for a nontrivial time 
dependence. Weyl semimetal can thus be thought of as being characterized by a $\theta$-term with the simplest nontrivial space and time 
dependence of the ``axion field" $\theta$. 

Integrating by parts and eliminating a total derivative term, we can rewrite Eq.~\eqref{eq:1} in the Chern-Simons form~\cite{Jackiw90}
\beq
\label{eq:33}
S_{\theta} = - \frac{e^2}{8 \pi^2} \int dt d \br \, \partial_{\mu} \theta \epsilon^{\mu \nu \alpha \beta} A_{\nu} \partial_{\alpha} A_{\beta}. 
\eeq
Varying Eq.~\eqref{eq:33} with respect to the vector potential, we obtain the following expression for the current
\beq
\label{eq:34}
j_{\nu} = \frac{e^2}{2 \pi^2} b_{\mu} \epsilon^{\mu \nu \alpha \beta} \partial_{\alpha} A_{\beta}, \,\,\, \mu = 1, 2, 3, 
\eeq
and 
\beq
\label{eq:35}
j_{\nu} = - \frac{e^2}{2 \pi^2} b_0 \epsilon^{0 \nu \alpha \beta} \partial_{\alpha} A_{\beta}. 
\eeq
It is easy to see that Eq.~\eqref{eq:34} represents the anomalous Hall effect,~\cite{Burkov11-1}
while Eq.~\eqref{eq:35} the chiral magnetic effect,~\cite{Kharzeev08,Burkov12-1} i.e. generation of equilibrium
current in response to an applied magnetic field.  

It is worth noting that the chiral magnetic effect and the anomalous Hall effect are closely related to the topological magnetoelectric effect, 
characterizing TR-invariant topological insulators.~\cite{Qi08}
Indeed, we can rewrite Eq.~\eqref{eq:35} as
\beq
\label{eq:35.1}
\bj = - \frac{e^2}{4 \pi^2} \partial_{t} \theta \,\bB. 
\eeq
Using $\bj = \partial_{t} \bP$, where $\bP$ is the electric polarization, i.e. identifying $\bj$ with the polarization current (one of two 
types of currents, which may exist in the bulk of an insulator), we have
\beq
\label{eq:35.2}
\partial_{t} \bP = - \frac{e^2}{4 \pi^2} \partial_{t} \theta \, \bB, 
\eeq
which gives
\beq
\label{eq:35.3}
\bP = - \frac{e^2}{4 \pi} \bB, 
\eeq
in the TR-invariant case, when $\theta = \pi$, which is precisely the quantized topological magnetoelectric effect.~\cite{Qi08}
In a Weyl semimetal sample the chiral magnetic effect can also be measured as charge polarization (voltage), arising in response to an 
applied external magnetic field (the voltage will of course not be universal in this case). 

The anomalous Hall effect can similarly be related to another, equivalent, form for the topological magnetoelectric effect. 
Indeed, Eq.~\eqref{eq:34} can be written as
\beq
\label{eq:35.4}
\bj = \frac{e^2}{4 \pi^2} \bnabla \theta \times \bE. 
\eeq
Identifying $\bj$ with the magnetization current, $\bj = \bnabla~\times~\bM$ (second kind of current, possible in the bulk of an insulator), we obtain
\beq
\label{eq:35.5}
\bM = \frac{e^2}{4 \pi} \bE, 
\eeq
which is an equivalent form of the topological magnetoelectric effect. 
\section{Effect of spectral gap}
It is often stated that Weyl semimetal is a topologically stable phase, but only provided translational symmetry is preserved. 
Indeed, any potential (even random, but with a nonvanishing mean value), that can scatter electrons between the Weyl nodes, 
will open up a gap and eliminate the nodes. Even more alarmingly, in the presence of a nonvanishing chiral chemical potential $b_0$, 
which shifts the left and right nodes in opposite directions in energy and thus creates perfectly nested electron and hole Fermi surfaces, 
the translational symmetry will be broken spontaneously due to the formation of an excitonic condensate for arbitrarily weak electron-electron 
interactions. Thus, in this case, Weyl semimetal, strictly speaking, is never a ground state, and exists only at temperatures above the excitonic 
condensation transition temperature (which is most likely very low). 
In this section we show, that topological transport properties of the Weyl semimetal, which, as we have demonstrated in the previous section, 
are closely related to the chiral anomaly, in fact survive even when a spectral gap is opened due to either an external potential, or spontaneously, 
as a result of electron-electron interactions, provided the gap is small enough.  

We will focus on the case of the spectral gap resulting from the Coulomb interaction-driven formation of an excitonic condensate in the 
presence of nonzero chiral chemical potential (the final result should not depend on the origin of the gap). 
Let us first briefly demonstrate that nonzero chiral chemical potential $b_0$ leads to the spontaneous breaking of translational symmetry 
in the presence of electron-electron interactions. 

Adding electron-electron interactions to the Hamiltonian Eq.~\eqref{eq:6}, restricting ourselves to  
the lowest-energy degrees of freedom near the nested electron and hole Fermi surfaces, enclosing the 
right (R) and left (L) Weyl nodes correspondingly, and eliminating the $\bb \cdot \bsigma$ term by the chiral gauge transformation, 
discussed above, we obtain a BCS-like effective Hamiltonian
\beqa
\label{eq:36}
H&=&\sum_{\bk} \left[ (k - b_0) c^\dg_{\bk R} c^\pdg_{\bk R} + (-k + b_0) c^\dg_{\bk L} c^\pdg_{\bk L}\right] \nonumber \\
&+&\frac{U}{V} \sum_{\bk, \bk'} c^\dg_{\bk R} c^\dg_{\bk' L} c^\pdg_{\bk L} c^\pdg_{\bk' R}. 
\eeqa
Here $U$ is the screened Coulomb interaction potential, whose approximate value can be estimated as
\beq
\label{eq:37}
U \lesssim \lim_{\bq \rightarrow 0} \frac{4 \pi e^2}{q^2 + 2 g(b_0) 4 \pi e^2} = \frac{1}{2 g(b_0)}, 
\eeq
where $g(\epsilon) = \epsilon^2/2 \pi^2$ is the density of states of a single Weyl cone and the factor of $2$ in the denominator in Eq.~\eqref{eq:37}
comes from the two Weyl nodes. 
Introducing excitonic order parameter
\beq
\label{eq:38}
\Delta = \frac{U}{V} \sum_{\bk} \langle c^\dg_{\bk R} c^\pdg_{\bk L} \rangle, 
\eeq
and decoupling the interaction term in Eq.~\eqref{eq:36} in the Hartree-Fock approximation, we 
obtain the standard BCS equation for $\Delta$
\beq
\label{eq:39}
1 = \frac{U g(b_0)}{2} \int_0^{\xi_c} d \xi \frac{\tanh(\sqrt{\xi^2 + \Delta^2}/ 2 T)}{\sqrt{\xi^2 + \Delta^2}}, 
\eeq 
where $\xi_c$ is a cutoff of the order of $\Delta_{S,D}$ in Eq.~\eqref{eq:3}. 
This gives the critical temperature of the excitonic condensation
\beq
\label{eq:40}
T_c \sim \xi_c e^{-2/U g(b_0)} \lesssim \xi_c e^{-4}, 
\eeq
which can, in principle, be quite significant. 

We will now demonstrate that the spectral gap, which opens as a result of spontaneous, as described above, or due to an external potential, translational 
symmetry breaking, does not affect the induced $\theta$-term in the action of the electromagnetic field and thus does not affect the topological response
of the Weyl semimetal. 
A note of caution is in order here. The above statement is of course only true provided the gap is much smaller than the high-energy cutoff scale $\xi_c$, 
i.e. the energy scale at which the Weyl node dispersion starts significantly deviating from linearity. Once the gap becomes comparable to $\xi_c$, 
our conclusions, based on a low-energy model of a Weyl semimetal with linear Weyl node dispersion, can no longer be expected to hold. 

We introduce a fluctuating field $\Delta(\br, \tau)$ in the imaginary time action of the Weyl semimetal
\beqa
\label{eq:41}
S&=&\int d \tau d \br \,\, \psi^\dg \left[ \partial_{\tau} + i e A_0 + b_0 \tau^z + \tau^z \bsigma \cdot \left( -i \bnabla + e \bA\right.\right. \nonumber \\
&+&\left.\left.\bb \tau^z \right) - \frac{1}{2} \Delta(\br, \tau) \tau^+ - \frac{1}{2} \Delta^*(\br, \tau) \tau^- \right] \psi^\pdg. \nonumber \\
\eeqa
$\Delta(\br, \tau)$ has the following form
\beq
\label{eq:42}
\Delta(\br, \tau) = \Delta_0(\br, \tau) e^{- 2 i \bb \cdot \br}, 
\eeq
where $\Delta_0(\br, \tau)$ is a slowly-varying envelope function. 
The field $\Delta(\br, \tau)$ arises from Hubbard-Stratonovich (HS) decoupling of the electron-electron interaction term in Eq.~\eqref{eq:36} (we will leave out 
the term, quadratic in $\Delta$, for brevity, but it is implicitly understood to be present).
Rewriting the action in relativistic notation, we obtain
\beqa
\label{eq:43}
S&=&i \int d^4 x \bar \psi \left[ \left(\slashed{D}_L + i \Delta^*\right) \frac{1 - \gamma^5}{2} \right. \nonumber \\
&+&\left.\left( \slashed{D}_R + i \Delta \right)\frac{1 + \gamma^5}{2}  \right] \psi,
\eeqa
where $\slashed{D}_{R,L} = \gamma^{\mu} [\partial_{\mu} + i e A_{\mu} \pm i b_{\mu} (1-s)]$. 

As before, we want to extract the term, proportional to $b_{\mu}$, from the action, by a sequence of infinitesimal chiral gauge transformations~\eqref{eq:14}.
Following the same procedure as before, we consider the operator
\beq
\label{eq:44}
{\cal D} = \left(\slashed{D}_L + i \Delta^*\right) \frac{1 - \gamma^5}{2} + \left( \slashed{D}_R + i \Delta \right)\frac{1 + \gamma^5}{2}. 
\eeq
This operator is clearly not hermitian. We can, however, construct two hermitian operators
\beqa
\label{eq:45}
{\cal D}^\dg {\cal D}^\pdg&=&\left[ \slashed{D}_L^2 +  i \gamma_{\mu} \partial_{\mu} \Delta^* - 
\Delta^* \gamma^{\mu} b_{\mu} (1-s) \right]\frac{1 - \gamma^5}{2} \nonumber \\
 &+&\left[ \slashed{D}_R^2 - i \gamma^{\mu} \partial_{\mu} \Delta - \Delta \gamma^{\mu} b_{\mu} (1-s) \right] \frac{1 + \gamma^5}{2} + \Delta^* \Delta, \nonumber \\
\eeqa
and 
\beqa
\label{eq:46}
{\cal D}^\pdg {\cal D}^\dg&=&\left[ \slashed{D}_L^2 -  i \gamma_{\mu} \partial_{\mu} \Delta^* +
\Delta^* \gamma^{\mu} b_{\mu} (1 - s)\right]\frac{1 + \gamma^5}{2} \nonumber \\
 &+&\left[ \slashed{D}_R^2 - i \gamma^{\mu} \partial_{\mu} \Delta - \Delta \gamma^{\mu} b_{\mu} (1 - s)\right] \frac{1 - \gamma^5}{2} + \Delta^* \Delta, \nonumber \\
\eeqa
The two operators have identical nonnegative eigenvalues, but different eigenfunctions
\beqa
\label{eq:47}
{\cal D}^\dg {\cal D}^\pdg \phi_n(x)&=&\epsilon_n^2 \phi_n(x), \nonumber \\
{\cal D}^\pdg {\cal D}^\dg \tilde \phi_n(x)&=&\epsilon_n^2 \tilde \phi_n(x). 
\eeqa
We then expand the transformed Grassmann variables $\psi'(x)$ and $\bar \psi'(x)$ with respect to the complete sets $\phi(x)$ and $\tilde \phi(x)$ correspondingly
and obtain the following expression for the transformation Jacobian
\beq
\label{eq:48}
J = e^{i ds \,\int d^4 x \, \left[I(x) + \tilde I(x) \right] /2}, 
\eeq
where 
\beqa
\label{eq:49}
I(x)&=&\sum_n \phi^*_n(x) \theta(x) \gamma^5 \phi_n(x) \nonumber \\
&=&\lim_{M \rightarrow \infty} \sum_{n} \phi^*_n(x) \theta(x) \gamma^5 e^{-{\cal D}^\dg {\cal D}^\pdg/M^2} \phi_n(x), 
\eeqa
and 
\beqa
\label{eq:50}
\tilde I(x)&=&\sum_n \tilde \phi^*_n(x) \theta(x) \gamma^5 \tilde \phi_n(x) \nonumber \\
&=&\lim_{M \rightarrow \infty} \sum_{n} \tilde \phi^*_n(x) \theta(x) \gamma^5 e^{-{\cal D}^\pdg {\cal D}^\dg/M^2} \tilde \phi_n(x), 
\eeqa
Performing exactly the same manipulations as in Eqs.~\eqref{eq:26}-\eqref{eq:30}, it is then straightforward to show that the infinitesimal 
chiral transformation Jacobian has exactly the same form as in Eqs.~\eqref{eq:30}, \eqref{eq:31}, i.e. the spectral gap term in Eq.~\eqref{eq:43} 
does not contribute to the chiral-transformation-induced $\theta$-term in the action. 
The reason for this is easy to understand. The terms that can contribute to $I(x)$ and $\tilde I(x)$ in the limit $M \rightarrow \infty$ must 
be proportional to the product of four gamma-matrices, multiplied by the factor $1/M^4$, as all other terms get nullified by multiplying them with 
the $\gamma^5$ matrix and taking the trace. By examining Eqs.~\eqref{eq:45} and \eqref{eq:46} it is easy to convince oneself that such 
a term can only arise from the $\slashed{D}_{R,L}^2$ part, as only these contain two gamma matrices. 

The HS terms in Eq.~\eqref{eq:43} explicitly break the chiral symmetry of the action by mixing the left and right fermions. 
This means that the HS terms themselves change under the chiral transformation Eq.~\eqref{eq:14}. 
Indeed, the imaginary time action after the chiral transformation is given by
\beqa
\label{eq:51}
&&S = S_{\theta} + i \int d^4 x \, \, \bar \psi \nonumber \\
&\times& \left[ \slashed{D} + i \Delta_0^* e^{-2 i b_4 x_4} \frac{1 - \gamma^5}{2} + i \Delta_0 e^{2 i  b_4 x_4} \frac{1 + \gamma^5}{2} \right] \psi, \nonumber \\
\eeqa
where $\slashed{D} = \gamma^{\mu}(\partial_{\mu} + i e A_{\mu})$.  
Thus, in principle, while the dependence on $\bb$ has indeed been eliminated from the fermionic part of the action, the dependence on $b_4$ (i.e. $b_0$) remains, and could 
contribute to the electromagnetic part of the action after fermions are integrated out. 
It is easy to see, however, that this does not happen. Indeed, focusing on the fermionic part of the action in Eq.~\eqref{eq:51}, and integrating out fermions, we simply obtain the Ginzburg-Landau action for the excitonic 
order parameter $\Delta_0$, which must, based on symmetry and gauge invariance considerations, have the following general form
\beq
\label{eq:51.1}
S_{GL} =  \int d^4 x \left[\varrho \,\partial_{\mu}\Delta_0^* \partial_{\mu} \Delta_0 + r |\Delta_0|^2  + u |\Delta_0|^4 \right],
\eeq
where the coefficients $\varrho$, $r$, and $u$ depend on the gauge potential only through the field invariant  $F_{\mu \nu} F^{\mu \nu}$, since the 
excitonic order parameter $\Delta_0$ is charge-neutral, and $r \sim (T - T_c)/T_c$, with $T_c$ given by Eq.~\eqref{eq:40}. 
This clearly implies that the $\theta$-term in~\eqref{eq:51.1} is not renormalized by the excitonic part of the action. 
Physically this happens due to the fact that the gap-opening due to $\Delta$ is a low-energy phenomenon, while the $\theta$-term contains the
contribution of all filled states and thus can not be affected by $\Delta$.  The only way $\Delta$ could affect the $\theta$-term is through the chiral anomaly, 
which, as we have demonstrated above, also does not happen. 

\section{Discussion and conclusions}
We will now provide a less formal explanation of the above results, which helps to understand the physical origin of the insensitivity of topological response in Weyl semimetals to opening up a spectral gap 
due to broken translational symmetry. 
Consider the low-energy Weyl semimetal Hamiltonian Eq.~\eqref{eq:6}, to which we will 
add the node-mixing potential term later.  
We assume, for concreteness, that the vector $\bb$ is along the $\hat z$-direction, $\bb = b \hat z$. 
We will also assume that an external orbital magnetic field $\bB$ is applied to the system, along the $\hat z$-direction as well, $\bB = B \hat z$. 
For clarity of the presentation we will separately consider two cases: $b \neq 0,\, b_0 = 0$ and the general case $b \neq 0, \, b_0 \neq 0$. 
\begin{figure}[t]
\includegraphics[width=8cm]{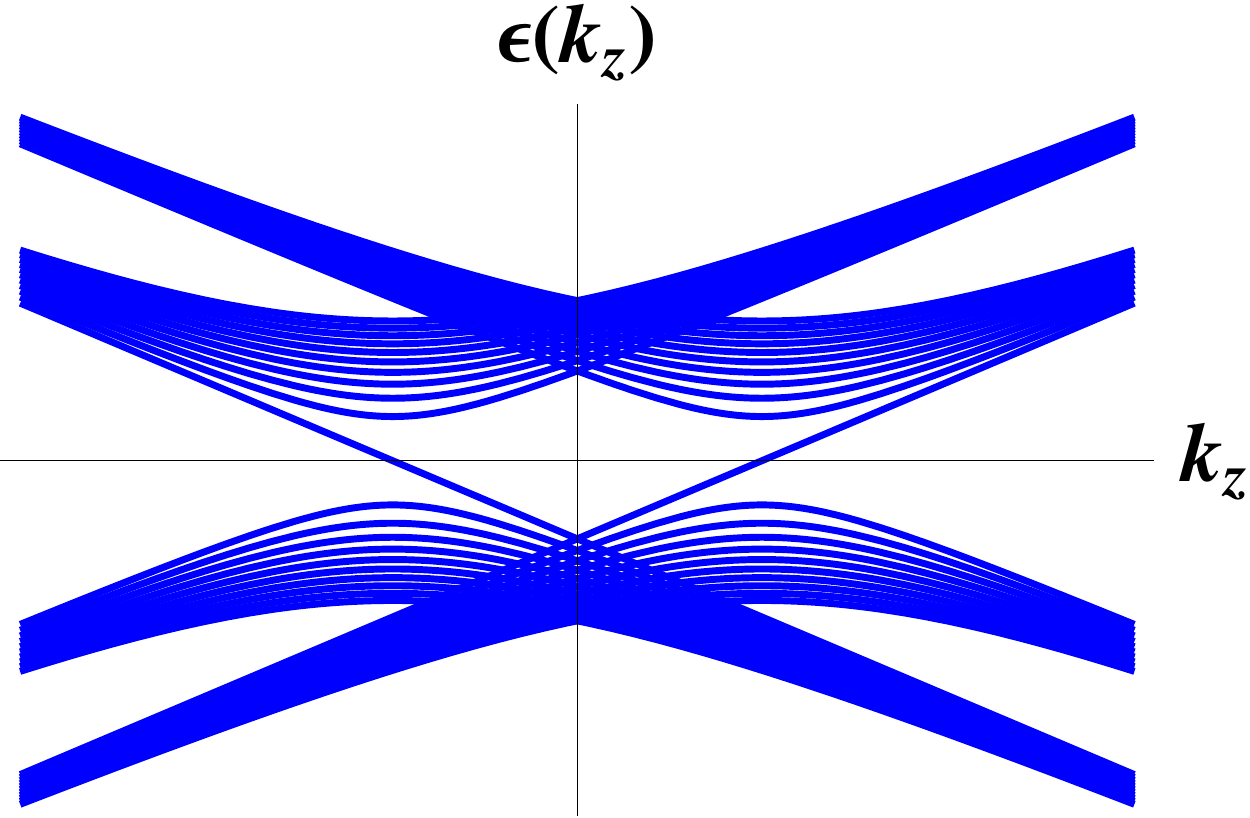}
\includegraphics[width=8cm]{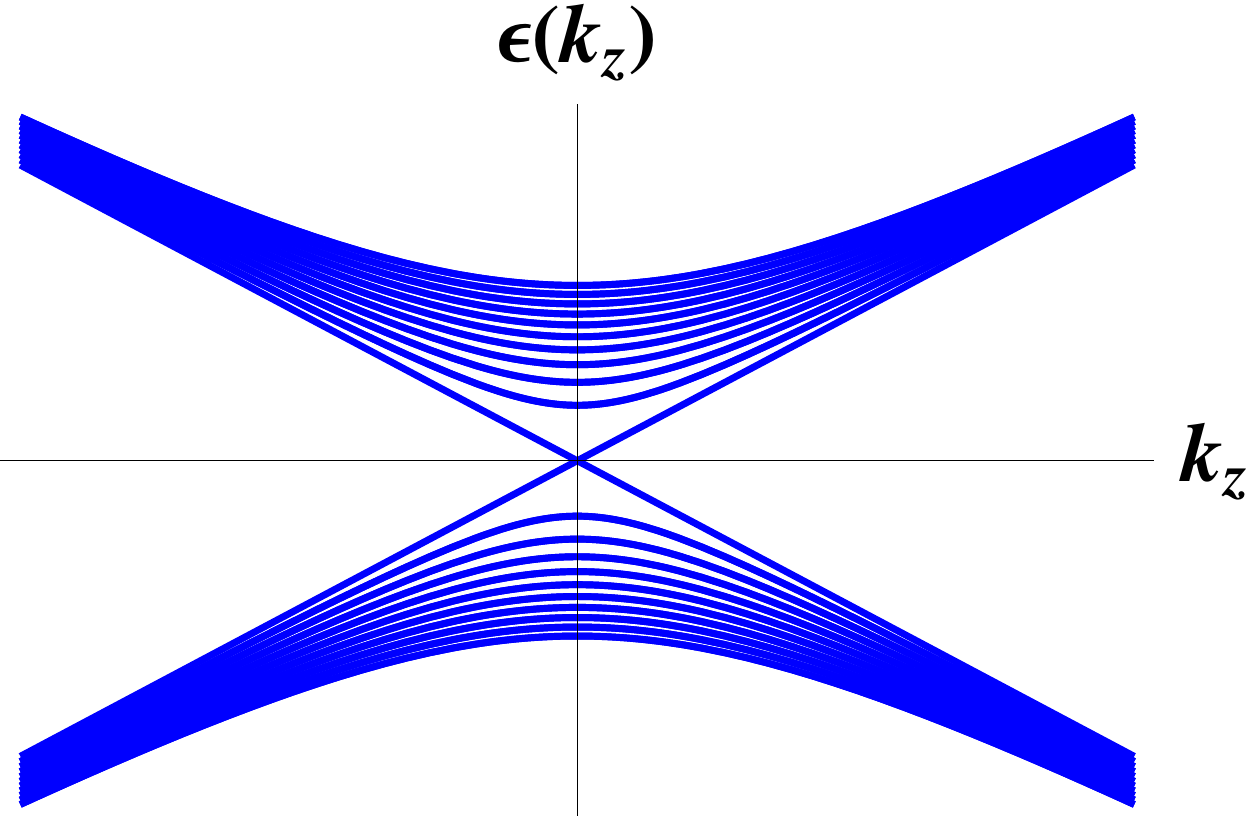}
\caption{(Color online) Landau level dispersion for $b \neq 0$ (top) and $b = 0$ (bottom) cases. It is clear that in the $b \neq 0$ case there is an
extra field-dependent electron density in the $n = 0$ Landau level, given by Eq.~\eqref{eq:56}.} 
\label{fig:1}
\end{figure}   
In the first case we obtain
\beq
\label{eq:52}
H = \tau^z (\sigma^x \pi_x + \sigma^y \pi_y) + \tau^z \sigma^z k_z + b \sigma^z, 
\eeq
where $\bpi = -i \bnabla + e \bA$ is the kinetic momentum in magnetic field. 
Introducing Landau level ladder operators $a = \ell_B (\pi_x - i \pi_y)/\sqrt{2}$ and  $a^\dg = \ell_B (\pi_x + i \pi_y)/\sqrt{2}$, where $\ell_B = 1/ \sqrt{e B}$ is the magnetic length, we obtain
\beq
\label{eq:53}
H = \frac{\omega_B}{\sqrt{2}} \tau^z (\sigma^+ a + \sigma^- a^\dg) + \sigma^z (b + \tau^z k_z), 
\eeq
where $\omega_B = 1/ \ell_B$. 
This is easily diagonalized and we obtain the Landau level dispersion as
\beq
\label{eq:54}
\epsilon_{n s \alpha} = s \sqrt{2 \omega_B^2 n + (\alpha k_z + b)^2}, \, n \geq 1, 
\eeq
where $s, \alpha = \pm$, while the $n = 0$ Landau level dispersions are given by
\beq
\label{eq:55} 
\epsilon_{n \alpha} =  - (\alpha k_z + b). 
\eeq
As obvious from Eq.~\eqref{eq:53}, the $n = 0$ Landau levels are polarized downwards. 
Since the $n \geq 1$ Landau levels are particle-hole symmetric, they do not contribute to the anomalous 
Hall conductivity. The contribution of the $n  = 0$ levels can be deduced using St\v{r}eda formula.~\cite{Streda82}
We note that the $n = 0$ Landau levels give an extra field-dependent electron density, compared to the $b = 0$ case (see Fig.~\ref{fig:1}), given by
\beq
\label{eq:56}
\delta n(B) = \frac{2 b}{2 \pi} \frac{1}{2 \pi \ell_B^2} = \frac{e b}{2 \pi^2} B. 
\eeq
The anomalous Hall conductivity can then be calculated as
\beq
\label{eq:57}
\sigma_{xy} = \lim_{B \rightarrow 0} e \frac{\partial \delta n(B)}{\partial B} = \frac{e^2 b}{2 \pi^2}. 
\eeq
Now suppose we turn on a weak periodic potential at wavevector $Q_z = 2 b$, hybridizing the Weyl nodes. 
In this case we need to fold Landau level dispersion to the reduced first BZ $- b \leq k_z < b$. 
Focusing on the $n = 0$ levels, the effect of the periodic potential will be to open a gap at the BZ boundary. It is clear, however, 
that the Hall conductivity remains unchanged, since $\delta n(B)$  remains unchanged after folding into the reduced BZ and gap opening. 
The expression for the Hall conductivity, Eq.~\eqref{eq:57}, can now be interpreted as conductance quantum $e^2/h$ per 
period $\pi/b$ of the Weyl node-hybridizing potential. 

Let us now consider the general case $b \neq 0,\, b_0 \neq 0$. In this case we can remove the term $b \sigma^z$ from the Hamiltonian by the chiral 
gauge transformation~\eqref{eq:14}, which gives rise to the corresponding $\theta$-term in the action, and focus on the effect of the $b_0 \tau^z$ term. 
In the presence of the magnetic field in the $\hat z$-direction and a uniform time-independent node-hybridizing potential $\Delta$, the Hamiltonian 
is given by
\beq
\label{eq:58}
H = \frac{\omega_B}{\sqrt{2}} \tau^z (\sigma^+ a + \sigma^- a^\dg) + \tau^z \sigma^z k_z + \tau^z b_0 - \Delta \tau^x. 
\eeq
Diagonalizing this we obtain the following Landau level dispersions
\beq
\label{eq:59}
\epsilon_{n s \alpha} = s \sqrt{\left(\sqrt{2 \omega_B^2 n + k_z^2} + \alpha b_0\right)^2 + \Delta^2},\,\, n \geq 1, 
\eeq
while the $n = 0$ Landau level dispersions are given by
\beq
\label{eq:60}
\epsilon_{0 \alpha} = \alpha \sqrt{(k_z - b_0)^2 + \Delta^2}, 
\eeq
where $s, \alpha = \pm$, as before. 
We can now calculate the current in response to the applied magnetic field (chiral magnetic effect)~\cite{Kharzeev08}
\beq
\label{eq:61}
j_ z = - \frac{e}{2 \pi \ell_B^2} \int_{-\Lambda}^{\Lambda} \frac{d k_z}{2 \pi} \frac{d}{d k_z}\left( \epsilon_{0 -} + \sum_{n = 1}^{\infty} \sum_{\alpha = \pm} \epsilon_{n - \alpha} \right), 
\eeq
where $d \epsilon_{n} / d k_z$ is the $\hat z$-component of the electron velocity in the $n$-th Landau level and $\Lambda$ is a cutoff momentum, which we will take to infinity at the end. 
Since $\epsilon_{n s \alpha}$ for $n \geq 1$ are even functions of $k_z$, only the $n = 0$ Landau level actually contributes to $j_z$. Then we obtain 
\beqa
\label{eq:62}
j_z&=&- \frac{e}{2 \pi \ell_B^2} \int_{-\Lambda}^{\Lambda} \frac{d \epsilon_{0 -}}{d k_z} = - \frac{e^2 B}{4 \pi^2} \left[ \epsilon_{0 - }(\Lambda) - \epsilon_{0 -}(- \Lambda) \right] \nonumber \\
&=&\frac{e^2 B}{4 \pi^2} \left( \Lambda - b_0 - \Lambda - b_0 \right) = - \frac{e^2 b_0}{2 \pi^2} B, 
\eeqa
where the second line is true in the limit $\Lambda/ \Delta \rightarrow \infty$. 
This coincides with Eq.~\eqref{eq:35}. 
Note that, in agreement with our previous discussion, the large-momentum states are important for this effect, which makes it insensitive to low-energy phenomena, such 
as the presence of the gap~$\Delta$.

In conclusion, in this work we have explicitly demonstrated that topological transport phenomena in Weyl semimetals are distinct manifestations of a single underlying phenomenon, 
the chiral anomaly, and are described  by a $\theta$-term in the action for the electromagnetic field, given by Eq.~\eqref{eq:1}. 
We have demonstrated that the $\theta$-term is insensitive to opening a gap in the spectrum of the Weyl semimetal due to broken (either by an external potential or spontaneously) 
translational symmetry, provided the gap is sufficiently small. 
 
\begin{acknowledgments}
We acknowledge financial support from the NSERC of Canada and a University of Waterloo start-up grant. 
\end{acknowledgments}


\begin{thebibliography}{99}
\bibitem{Vishwanath11} X. Wan, A.M. Turner, A. Vishwanath, and S.Y. Savrasov, Phys. Rev. B {\bf 83}, 205101 (2011). 
\bibitem{Balents11} L. Balents, Physics {\bf 4}, 36 (2011). 
\bibitem{Ran11} K.-Y. Yang, Y.-M. Lu, and Y. Ran, Phys. Rev. B {\bf 84}, 075129 (2011).
\bibitem{Burkov11-1} A.A. Burkov and L. Balents, Phys. Rev. Lett. {\bf 107}, 127205 (2011). 
\bibitem{Burkov11-2} A.A. Burkov, M.D. Hook, and L. Balents, Phys. Rev. B {\bf 84}, 235126 (2011).
\bibitem{Burkov12-1} A.A. Zyuzin, S. Wu, and A.A. Burkov, Phys. Rev. B {\bf 85}, 165110 (2012). 
\bibitem{Kim11} W. Witczak-Krempa and Y.-B. Kim, Phys. Rev. B {\bf 85}, 045124 (2012). 
\bibitem{Fang11}  G. Xu, H. Weng, Z. Wang, X. Dai, and Z. Fang, Phys. Rev. Lett. {\bf 107}, 186806 (2011).  
\bibitem{Halasz11} G.B. Hal\'{a}sz and L. Balents, Phys. Rev. B {\bf 85}, 035103 (2012). 
\bibitem{Hosur11} P. Hosur, S.A. Parameswaran, and A. Vishwanath, Phys. Rev. Lett. {\bf 108}, 046602 (2012). 
\bibitem{Aji11} V. Aji, Phys. Rev. B {\bf 85}, 241101 (2012). 
\bibitem{Carpentier12} P. Delplace, J. Li, and D. Carpentier, Europhys. Lett. {\bf 97}, 67004 (2012). 
\bibitem{Son12} D.T. Son and N. Yamamoto, arXiv:1203.2697 (unpublished). 
\bibitem{Qi12} C.-X. Liu, P. Ye, and X.-L. Qi, arXiv:1204.6551 (unpublished). 
\bibitem{Grushin12} A.G. Grushin, Phys. Rev. D {\bf 86}, 045001 (2012). 
\bibitem{Balents12} T. Meng and L. Balents, arXiv:1205.5202 (unpublished). 
\bibitem{Kolomeisky12} E.B. Kolomeisky and J.P. Straley, arXiv:1205.6354 (unpublished). 
\bibitem{Garate12} I. Garate and L. Glazman, Phys. Rev. B {\bf 86}, 035422 (2012). 
\bibitem{Jiang12} J.-H. Jiang, Phys. Rev. A {\bf 85}, 033640 (2012). 
\bibitem{Volovik03} G.E. Volovik, {\em The Universe in a Helium Droplet} (Clarendon Press, Oxford, 2003). 
\bibitem{Volovik05} F.R. Klinkhamer and G.E. Volovik, Int. J. Mod. Phys. A {\bf 20}, 2795 (2005).
\bibitem{Volovik07} G.E. Volovik, Lect. Notes Phys. {\bf 718}, 31 (2007).
\bibitem{Volovik11} G.E. Volovik, arXiv:1111.4627 (unpublished). 
\bibitem{Murakami} S. Murakami, New J. Phys. {\bf 9}, 356 (2007). 
\bibitem{Adler69} S. Adler, Phys. Rev. {\bf 177}, 2426 (1969). 
\bibitem{Jackiw69} J.S. Bell and R. Jackiw, Nuovo Cimento {\bf 60A}, 4 (1969).  
\bibitem{Nielsen83} H.B. Nielsen and M. Ninomiya, Phys. Lett. {\bf 130B}, 389 (1983). 
\bibitem{Fujikawa} K. Fujikawa and H. Suzuki, {\em Path Integrals and Quantum Anomalies} (Clarendon Press, Oxford, 2004). 
\bibitem{Nakahara} M. Nakahara, {\em Geometry, Topology and Physics} (Institute of Physics Publishing, Bristol and Philadelphia, 2002). 
\bibitem{Jackiw84} R. Jackiw, Phys. Rev. D {\bf 29}, 2375 (1984). 
\bibitem{Fradkin86}E. Fradkin, E. Dagotto, and D. Boyanovsky, Phys. Rev. Lett. {\bf 57}, 2967 (1986). 
\bibitem{Semenoff84} G. Semenoff, Phys. Rev. Lett. {\bf 53}, 2449 (1984). 
\bibitem{Haldane88} F.D.M. Haldane, Phys. Rev. Lett. {\bf 61}, 2015 (1988). 
\bibitem{Ludwig94} A.W.W. Ludwig, M.P.A. Fisher, R. Shankar, and G. Grinstein, Phys. Rev. B {\bf 50}, 7526 (1994). 
\bibitem{Kane} M.Z. Hasan and C.L. Kane, Rev. Mod. Phys. {\bf 82}, 3045 (2010). 
\bibitem{Zhang} X.-L. Qi and S.-C. Zhang, Rev. Mod. Phys. {\bf 83}, 1057 (2011). 
\bibitem{Nielsen81} H.B. Nielsen and N. Ninomiya, Nucl. Phys. {\bf B185}, 20 (1981); {\it ibid.} {\bf B193}, 173 (1981).
\bibitem{Vilenkin80} A. Vilenkin, Phys. Rev. D {\bf 22}, 3080 (1980). 
\bibitem{Cheianov98} A.Y. Alekseev, V.V. Cheianov, and J. Fr\"ohlich, Phys. Rev. Lett. {\bf 81}, 3503 (1998).   
\bibitem{Kharzeev08} K. Fukushima, D.E. Kharzeev, and H.J. Warringa, Phys. Rev. D {\bf 78}, 074033 (2008).  
\bibitem{Kharzeev11} D.E. Kharzeev and H.-U. Yee, Phys. Rev. D {\bf 83}, 085007 (2011). 
\bibitem{Abelev09} B.I. Abelev {\it et al.}, Phys. Rev. Lett. {\bf 103}, 251601 (2009). 
\bibitem{Wilczek87} F. Wilczek, Phys. Rev. Lett. {\bf 58}, 1799 (1987). 
\bibitem{Fujikawa79} K. Fujikawa, Phys. Rev. Lett. {\bf 42}, 1195 (1979). 
\bibitem{Vishwanath10} P. Hosur, S. Ryu, and A. Vishwanath, Phys. Rev. B {\bf 81}, 045120 (2010). 
\bibitem{Franz10} M.M. Vazifeh and M. Franz, Phys. Rev. B {\bf 82}, 233103 (2010). 
\bibitem{Jackiw90} S.M. Carroll, G.B. Field, and R. Jackiw, Phys. Rev. D {\bf 41}, 1231 (1990).   
\bibitem{Qi08} X.-L. Qi, T.L. Hughes, and S.-C. Zhang, Phys. Rev. B {\bf 78}, 195424 (2008). 
\bibitem{Streda82} P. St\v{r}eda, J. Phys. C {\bf 15}, L717 (1982). 
\end{thebibliography}
\end{document}